\documentstyle[pre,aps,psfig,epsf,epsfig,twocolumn]{revtex}

\tighten

\begin{document}

\title{\bf Equilibrium crystal shapes in the Potts model}
\author{R.P. Bikker, G.T. Barkema and H. van Beijeren}
\address{Institute for Theoretical Physics, Utrecht University,
Princetonplein 5, 3584 CC Utrecht, The Netherlands}
\date{\today}

\maketitle

\begin{abstract}
The three-dimensional $q$-state Potts model, forced into coexistence by fixing
the density of one state, is studied for $q=2$, 3, 4, and 6.  As a function of
temperature and number of states, we studied the resulting equilibrium droplet
shapes.  A theoretical discussion is given of the interface properties at
large values of $q$.  We found a roughening transition for each of the
numbers of states we studied, at temperatures that decrease with increasing
$q$, but increase when measured as a fraction of the melting temperature.  We
also found equilibrium shapes closely approaching a sphere near the melting
point, even though the three-dimensional Potts model with three or more states
does not have a phase transition with a diverging length scale at the melting
point. 
\end{abstract}

\section{Introduction}

If a binary mixture of fixed composition is brought into a coexistence
state it will phase-separate into two pure phases, separated by
an interface with a shape that minimizes the free energy of the
system~\cite{newman}. The phase with the smaller volume will typically
organize itself into a compact shape, known as the equilibrium shape. If
the surface tension (excess free energy per unit interface area) is
isotropic, this equilibrium shape will be a sphere. If either of the
co-existing phases is crystalline, the anisotropic surface tension will
lead to an aspheric equilibrium crystal shape (ECS). The ECS and the
orientation dependence of the surface tension are intimately related. Once
the surface tension is known for all orientations, the so-called Wulff
construction~\cite{Wulff,RW} allows the generation of the ECS. The reverse
is also possible, i.e., determining the surface tension as a function of
orientation from the ECS; this is a procedure followed in this manuscript.

The ECS has been the subject of experimental 
studies~\cite{HM,MH,BMH,Pav,Pav2,BW,SAC,ABW}.
One striking feature the ECS may show is a roughening transition: the
disappearance of facets (macroscopically flat surfaces) under rising
temperature. In practice equilibrium shapes are rarely seen and they are
hard to produce experimentally, but this does not mean they are without
practical importance.  For example the presence or absence of facets
in the ECS influences growth properties, such as the speed of growth
and the growth mode, even though growth shapes may differ strongly from
the ECS.  In general, the determination of the orientation dependence
of the surface tension is a tough problem. Experimentally it is very
difficult to measure, mainly due to equilibration problems. Numerical
studies suffer from the same equilibration problems, and in the most
common simulation techniques each surface orientation requires a separate
simulation. Theoretical results only have been obtained for simplified
models like the BCSOS model~\cite{BN}.

The prototypical model in which such properties are studied numerically,
is the conserved-order-parameter Ising model with nearest-neighbor
interactions, in which the total magnetization is kept constant. In the
two-dimensional Ising model, the ECS is a square at zero temperature,
but at any finite temperature the ECS looses all flat faces. It gradually
changes into a circle when the temperature is approaching the critical
temperature.  The behavior is richer in the three-dimensional Ising
model. At zero temperature, the ECS is a cube. At finite temperatures
below the roughening temperature $T_R$, the ECS still has macroscopically
flat faces but the corners as well as the edges are rounded. Above the
roughening temperature, the ECS does no longer feature macroscopically
flat faces. If the temperature is increased from the roughening
temperature to the critical temperature, the ECS gradually approaches
a sphere.

An extension of the Ising model is the Potts model, defined by
the Hamiltonian
\begin{equation}
H=-J\sum_{\langle i,j \rangle} \delta(\sigma_i,\sigma_j),
\end{equation}
in which $J$ is the coupling constant, $\delta$ denotes the Kronecker delta
function, and the summation runs over all pairs of nearest-neighbor sites,
each having a spin with value $\sigma=1\dots q$. Note that the two-state Potts
model is equivalent to the Ising model. The topic of this manuscript is to
study how the ECS and related properties such as the roughening temperature
behave with increasing number of states. Like in two dimensions the model
undergoes a phase transition at a temperature we will refer to as the melting
temperature $T_m$. Below this temperature the model has $q$ different phases,
each of which is dominated by one of the $q$ possible spin values, whereas for
temperatures above $T_m$ there is only a single phase in which on average all
of the spin values are present in equal amounts. It is known that for $q \geq
3$ the melting transition is of first order~\cite{barkboer}, in contrast to
the two-state (Ising) case where the transition is continuous.  It is
therefore not clear {\em a priori} whether there is a roughening transition
when $q \geq 3$, nor in how far the ECS should approach a sphere when $T$
approaches $T_R$. Furthermore, one should expect the surface tension to
approach a non-zero limit as the melting temperature is approached from below,
whereas for continuous transitions it vanishes in this limit.

The manuscript is organized as follows. First, in section~\ref{se:ECS} we
outline the numerical procedure to efficiently determine the ECS and show the
resulting shapes for all numbers of states for which accurate numerical
estimates of the melting temperature have been reported: $q=2$ (Ising), $3, 4$
and $6$. In section~\ref{se:largeq} we present some theoretical considerations
on the behavior of the model for large values of $q$. Section~\ref{se:Tr}
covers the measurements on the location of the roughening transition. In
section~\ref{se:surftens} we present the orientation-dependent surface tension
as extracted from the data on the ECS. We conclude with a discussion of the
results and an outlook to further research.

\section{Obtaining the ECS}
\label{se:ECS}

Equilibrium shapes can be determined numerically for the $q$-states Potts
model by forcing it into coexistence. This may be achieved by enforcing a
constraint on the spin densities. The richest behavior is observed in case 
the maximum of $q-1$ constraints are enforced, but mostly the resulting
configurations are still determined by the orientation dependence of the
surface tension, and can thus be obtained indirectly from numerical
simulations with a single constraint enforced.

The constraint enforced in our simulations is the conservation of the density
of state 1, while the densities of all other states are allowed to fluctuate
freely. To implement this, we combined a density-conserving cluster algorithm,
described in a recent article~\cite{bikbark}, with Glauber
dynamics~\cite{newman}, constrained to never flip spins into or out of state
1. We chose the ratio of cluster updates and Glauber updates per site to be
unity (i.e., a comparable amount of computational effort was spent to
each). The run-times of the simulations varied between 400\,000 and 800\,000
Monte Carlo updates per site and we typically allowed the system to thermalize
for 60\,000 to 100\,000 time steps. The lateral system size was 50 sites in
all lattice directions, with periodic boundary conditions, and the fraction of
conserved spins was typically $0.25$. In order to directly obtain the ECS we
repeatedly took "snapshots" of the evolving system, each time centering the
cluster around the origin. Taking the autocorrelation time into account, we
typically obtained 2000 independent equilibrium snapshots from one
simulation. At the end we calculated the time-averaged density profile of the
conserved component. Our results for the ECS were obtained from the 50\%
iso-density surface in this time-averaged density profile.

\begin{figure}
\begin{center}
\epsfxsize 4cm
\epsfbox{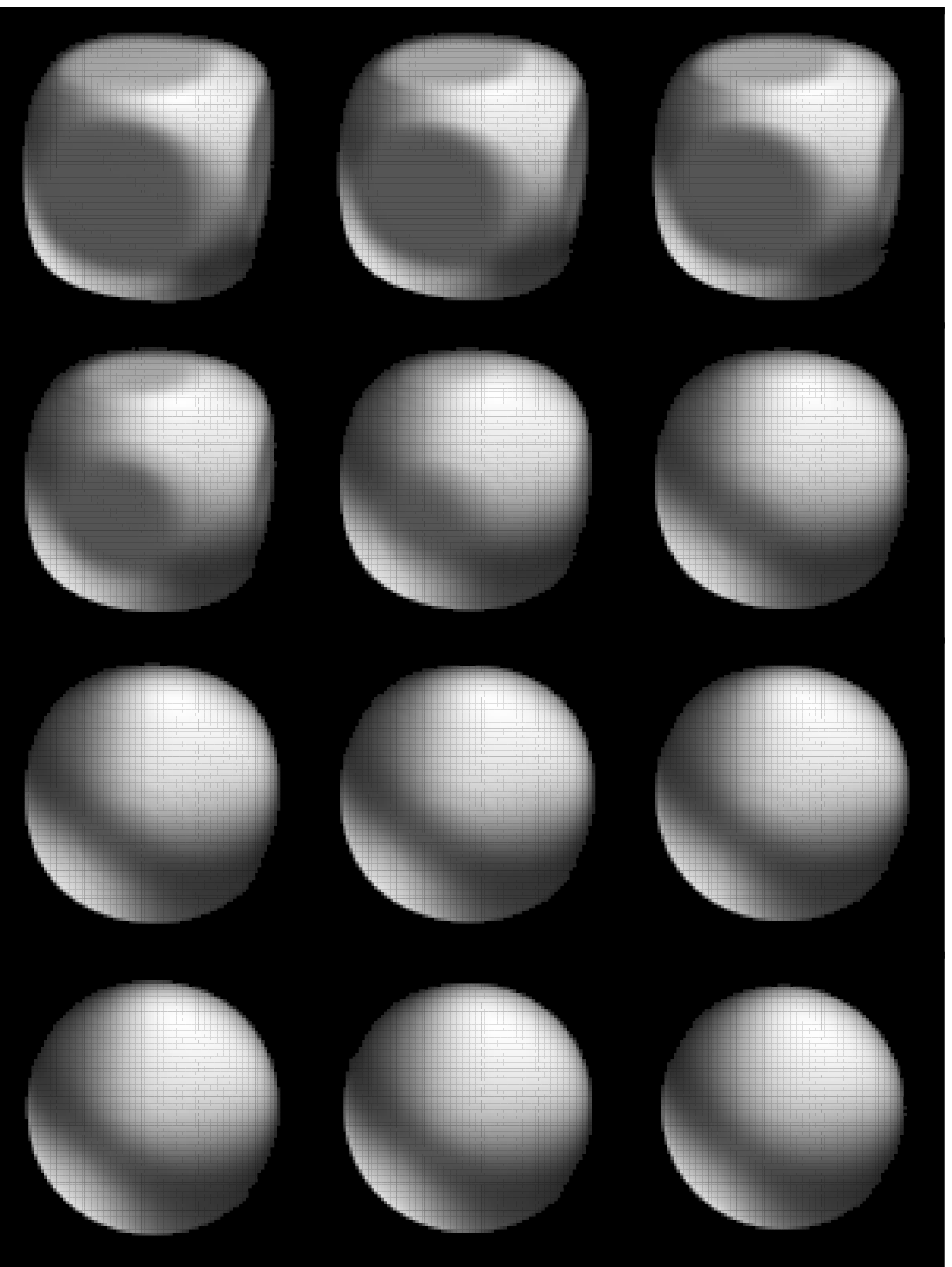}
\hspace{0.15cm}
\epsfxsize 4cm
\epsfbox{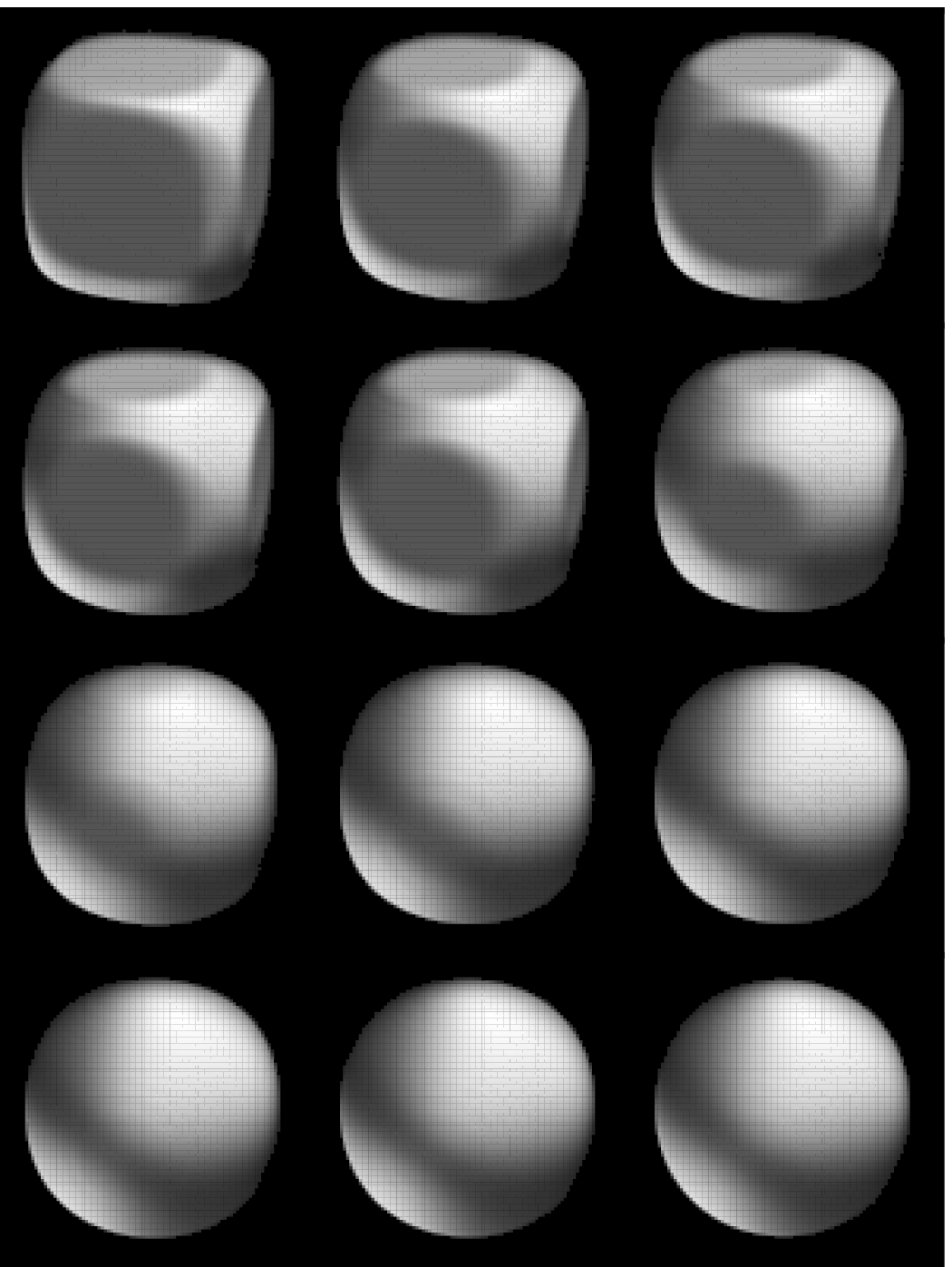}
\nopagebreak
\vspace{0.3cm}
\nopagebreak
\epsfxsize 4cm
\epsfbox{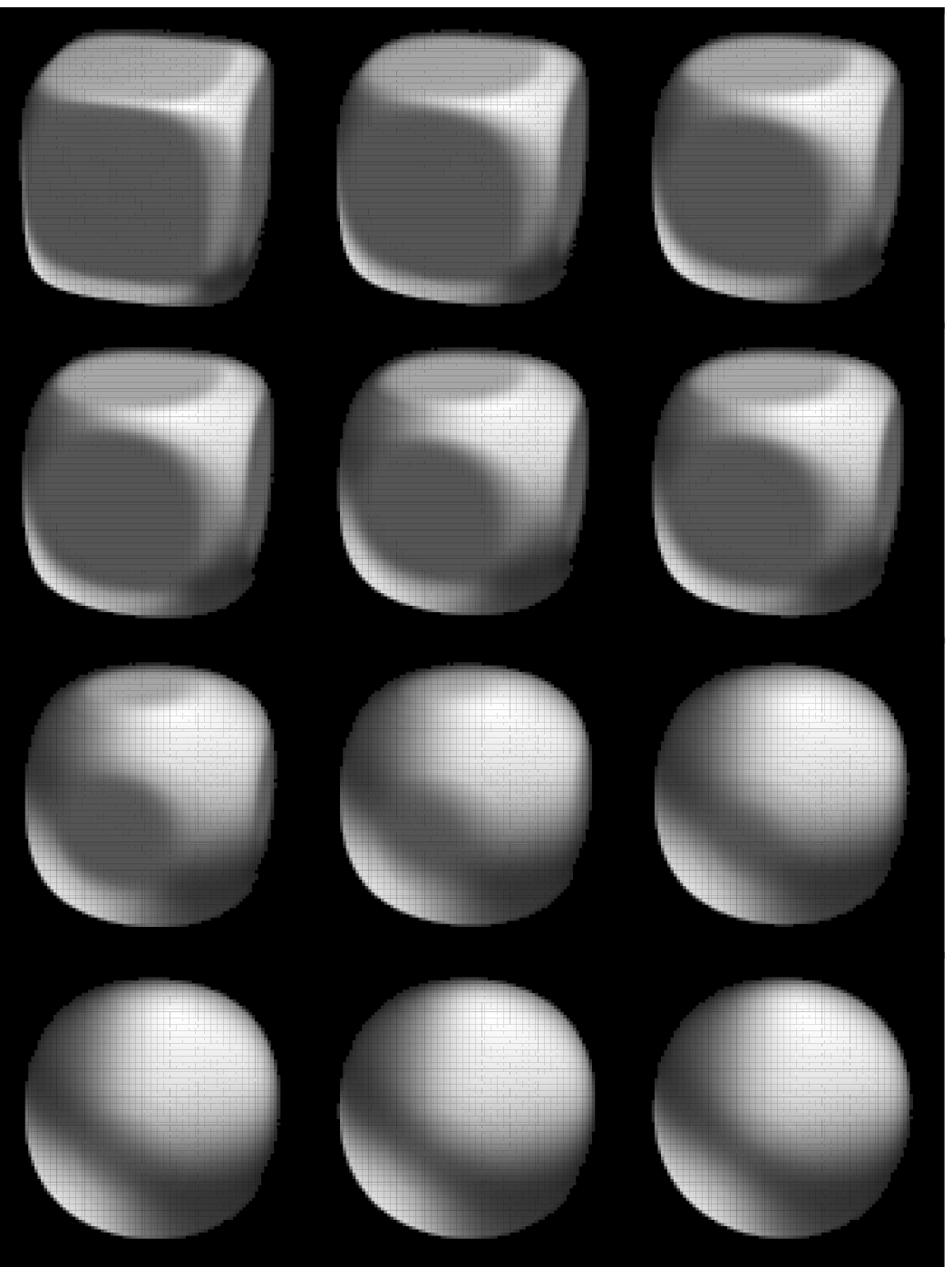}
\hspace{0.15cm}
\epsfxsize 4cm
\epsfbox{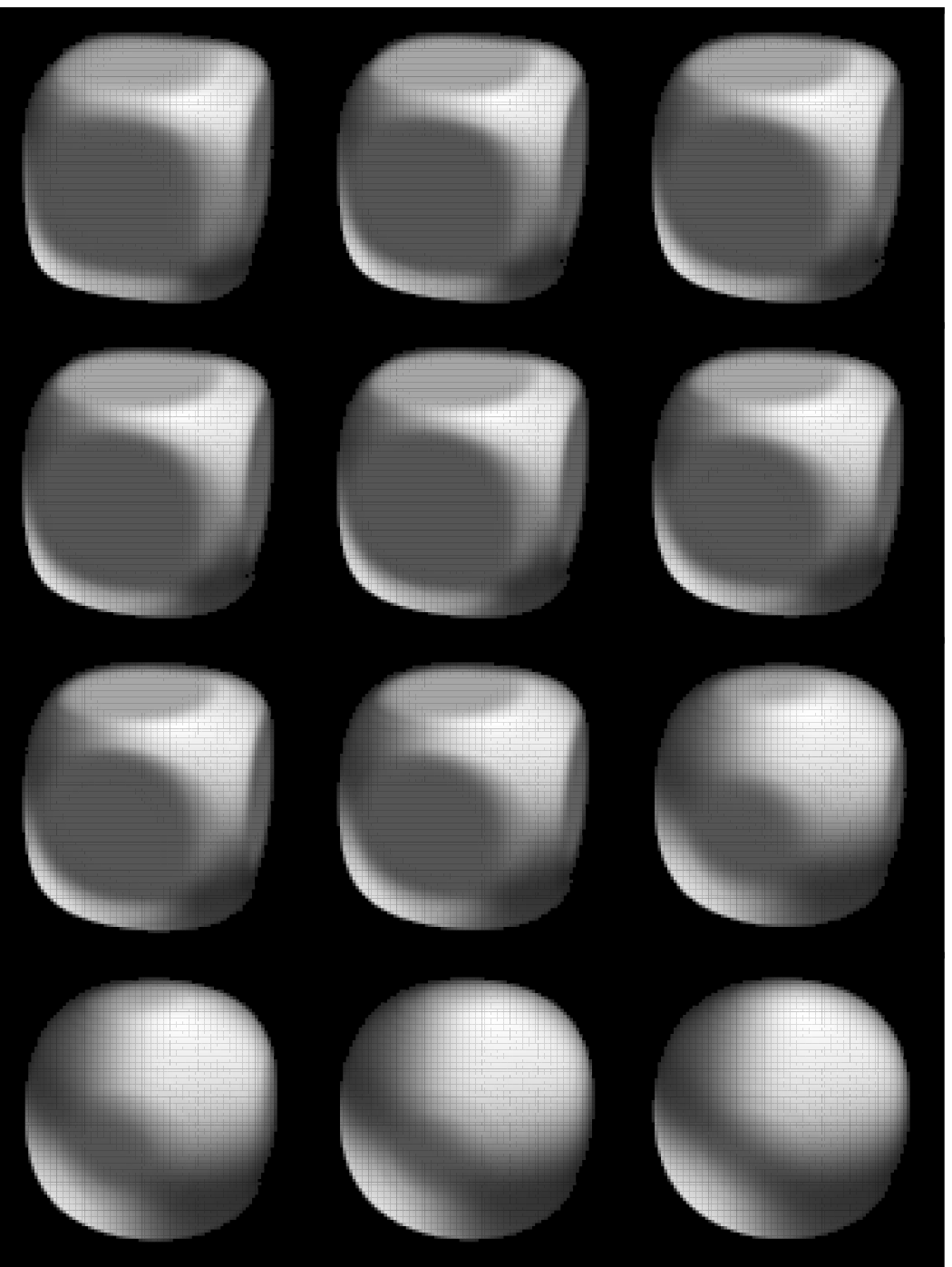}
\vspace{0.3cm}
\caption{\label{Shapes}
Symmetrized Equilibrium Cluster Shapes.  From top-left to bottom-right, the
boxes show the shapes for the 2, 3, 4 and 6-state Potts model,
successively. Within each box, again from top-left to bottom-right, the
temperature increases from 0.25 $T_m$ to 0.80 $T_m$, in increments of 0.05
$T_m$.}
\end{center}
\end{figure}

In figure~\ref{Shapes}, we show the equilibrium crystal shapes for
several temperatures, for the case of $q=2$ (Ising), $3$, $4$, and $6$
states, respectively. These shapes are obtained from the 50\% iso-density
surface in the density profile, after full symmetrization, i.e. after
we averaged over all 48 possible invariant mirror images of the cube.

The simulation temperatures are chosen as fractions of the estimates
for the melting temperatures reported in Refs.~\cite{BLH,Mach,Wu};
these values are given in table~\ref{taTr}.

\section{Behavior for large $\lowercase{q}$}
%
%
%
\label{se:largeq}

For large values of $q$ the thermal evolution of the ECS can be
understood from the following arguments.  First of all notice that at
low temperatures the $q$ phases of the $q$-state Potts model without
conservation laws will be very close to the zero temperature phases in
which all spins on the lattice have equal value. In $d$ dimensions the
free energy per site $f$ of these states therefore satisfies
\begin{equation} \label{tlow}
f(T_{\mathit{low}})\approx -dJ.
\end{equation}
The high temperature phase is completely dominated by the entropy
resulting from the $q$ different occupation options for each site,
with a resulting entropy per site of $k_B \ln q$ and free energy
\begin{equation} \label{thigh}
f(T_{\mathit{high}})\approx -k_B T \ln q.
\end{equation}
The melting temperature is obtained to a good approximation by equating
these two expressions, with the result
\begin{equation} \label{betaqlarge}
\beta_m J \approx \frac 1 d \ln q
\end{equation}
for large $q$, with $\beta=1/k_B T$. Note that for $d=2$ this is in
agreement with the exact result~\cite{Wu} $\beta_m J=\ln (1+\sqrt{q})$.
For large $q$ one sees that $J \gg k_B T$ for temperatures below $T_m$.
Therefore local excitations, typically consisting of single overturned
spins, will be extremely rare in the low temperature phases, as they
require an energy far exceeding the gain in free energy due to the entropy
increase. Similarly the tendency to form clusters of equal spin in the
high temperature phase is very weak, as the resulting gain of energy by
far does not compensate the resulting loss of entropic free energy. As
a result the melting transition for high $q$ will become very sharp.

Now we can also estimate the orientation dependence of the surface
tension. In good approximation, an interface of 001-orientation (or
symmetrically equivalent) is just a flat interface between two pure states,
with a free energy of $J$ (one broken bond) per unit area. Similarly, by
counting broken bonds, one finds that an interface of $kl1$-orientation
(here $k$ and $l$ are not necessarily integers) has an energy per unit area of
$(1+k+l)/\sqrt{1+k^2+l^2}$. For non-zero temperature there is an additional
contribution from the entropy of the steps required to form such an
interface. For not too large values of $k$ and $l$ this may be obtained by 
multiplying the step density needed to create a surface of orientation $kl1$ 
with the meandering entropy of such a step. The resulting expression is.
\begin{equation} \label{f(T)}
f(T)=\frac{(1+k+l)J-k_B T \ln\left[(k+l)^{k+l}/(k^k l^l)\right]}
{\sqrt{1+k^2+l^2}}.
\end{equation}
This result is independent of the value of $q$, especially it also
holds for the Ising model. Excitations creating spins not belonging
to the two phases that coexist at the interface are so rare that they
may be neglected completely.  For $k/l$ or $l/k \ll \exp(-\beta J)$
the entropy becomes dominated by thermal step fluctuations, but this
concerns an extremely small range of orientations only (However, it does
set the distance between facet edges). Bl\"ote and Hilhorst~\cite{BH} give 
expressions for the free energy from which the low temperature ECS may be
obtained for all orientations. 

The unimportance of all but the two coexisting phases implies that for large
$q$ the 001-facets do not roughen, as $T_m$ is well below the roughening
temperature of these facets in the Ising model (see Section~\ref{se:Tr}). The
equilibrium shape then remains nearly cubic for all temperatures up to $T_m$.

\section{Location of the roughening transition}
\label{se:Tr}

Roughening transitions of crystal surfaces are characterized by the
disappearance of macroscopically flat regions, or facets, in the
ECS. Facets existing at low temperatures often disappear when the
temperature increases, at a characteristic temperature known as the
roughening temperature $T_R$ of the specific facet orientation. At this
temperature the inverse radius of curvature $(R_c)^{-1}$ of the crystal
surface at the center of the facet jumps from zero (when a facet is
present) to a non-zero value.  Using renormalization-group calculations,
Jayaprakash {\em et al.}~\cite{JST,BN} showed that the size of this jump
satisfies the universal relationship
\begin{equation} \label{RC}
R_c = \frac{z_0kT_R\pi}{2\gamma_0},
\end{equation}
where $z_0$ is the distance from the tangent plane at the facet to the
center of mass of the crystal, and $\gamma_0$ is the surface tension
of the 001-facet at $T=T_R$. The latter is not known exactly for any
of the $q$-state Potts models in three dimensions. An approximation
consisting of the ground state value plus the first correction term in
a low temperature expansion is
\begin{equation} \label{G0}
\gamma_0 = \beta J - 2e^{-4\beta J}.
\end{equation}
It will turn out that this approximation suffices for our goal: even
for the smallest $q$-values the contribution of the correction term at
$T=T_R$ is already smaller than the estimated systematic error in the
curvature measurements.

If the center of mass of the ECS is placed at the origin, the centers
of the facets that are present below the roughening temperature are
located on the principal axes. To estimate the location of the roughening
transition, we first measure the curvature at the six points where the ECS
(obtained as in the previous section) cuts a principal axis, in the two
principal directions tangential to the surface. The curvature was obtained
by fitting a quadratic function $y = y_0 - \frac12 y_1 x^2$; the fitted
result for $y_1/y_0$ is the normalized inverse radius of curvature.
Since we use a non-zero fitting range, the slope of the curve cannot
become infinite and consequently we do not observe a jump in curvature;
in fact we find quite smooth dependence of curvature on temperature,
as a result of finite size, round-off and fluctuation effects. However,
we made sure that our procedure to estimate $T_R$ is not very sensitive
to this. Notably our estimate of the roughening temperature for the
Ising model is close to previous estimates: 
Adler reported
$T_R/T_m = 0.55 \pm 0.02$~\cite{adler}, Mon et al. reported
$T_R/T_m = 0.54 \pm 0.02$~\cite{mon}, and Holzer and Wortis reported
$T_R/T_m = 0.545 \pm 0.004$~\cite{holzer}.

\begin{figure}
\begin{center}
\epsfxsize 8cm \epsfbox{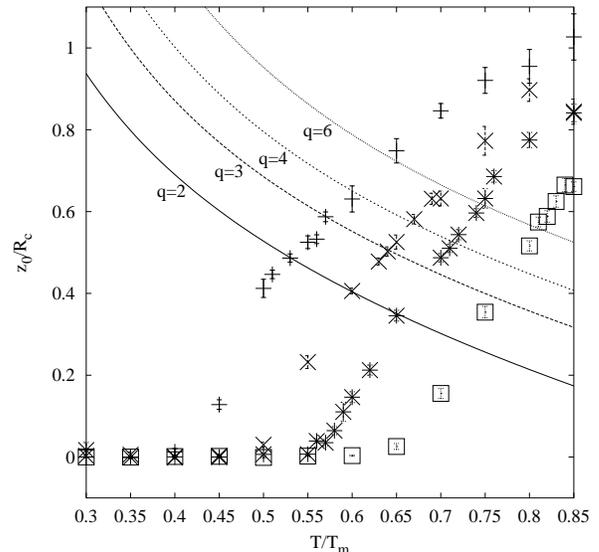}
\caption{\label{curves}
Normalized inverse curvature for the Potts model with $q=2$
(pluses), 3 (crosses), 4 (stars) and 6 states (squares). 
The intersection points of the lines, described by Eq.~(\ref{RC}), and the
data points are estimates for the roughening temperature $T_R$. }
\end{center}
\end{figure}

The resulting measurements of the normalized inverse curvature are plotted
in figure~\ref{curves}. The error bars indicate the statistical error;
additional simulations for some points indicate that the systematic errors
arising from the effects of thermalization, cluster shape deformations,
long lived thermal excitations, etc., are larger. In the same plot, the
curves described by Eq.~(\ref{RC}) are also plotted. The intersection
points of these two curves are estimates of the location of the roughening
transition. The resulting values for the roughening temperature can be
found in table~\ref{taTr}; the error indications are our estimates of
the statistical and systematic errors combined.

In the previous section we noted that for large $q$ there is no roughening
transition. Obviously $q=6$ is not large enough to observe this, but the
increase of $T_R/T_m$ is very clear.  It would be interesting to know the
largest $q$-value for which a stable roughening transition does exist. The
values of $\beta J_R$ quoted in table~\ref{taTr} combined with the estimate of
$T_m$ of Eq.~(\ref{betaqlarge}) suggest a value between 20 and 40. With the
present method this would make simulations very slow, and it would also
require an accurate measurement of the melting temperatures, since no
literature values exist.

\begin{table}[h] {\bf Roughening temperatures, $q$-state Potts-model}
\begin{tabular}{lllll}
$q$ & $\beta J_R$ & $\beta J_m$ & $T_R/T_m$ & $\beta J_m (2D)$ \\
\hline
$2$ \cite{BLH}  & $0.84 (2)$ & $0.443309 (2)$ & $0.53 (3)$ & 0.8813\\ 
$3$ \cite{Mach} & $0.86 (2)$ & $0.5505 (1)$   & $0.64 (3)$ & 1.0051\\ 
$4$ \cite{Wu}   & $0.89 (2)$ & $0.631 (2)$    & $0.71 (3)$ & 1.0986\\ 
$6$ \cite{Wu}   & $0.93 (2)$ & $0.751 (2)$    & $0.81 (3)$ & 1.2382
\end{tabular}
\caption{\label{taTr} Inverse transition temperature, ratio of the
roughening and transition temperatures and inverse roughening temperature
of the $q$-states Potts model. The references in the first column point
to the sources we used for the data on the transition temperatures. For
comparison we added the critical interaction parameter $\beta J_m =
\log(1+\sqrt{q})$ for the two dimensional Potts model in the last
column.}
\end{table}

\section{Orientation dependence of the surface tension}
\label{se:surftens}

In general, the surface tension $\gamma$ can be written as $\gamma=\gamma_0
\gamma(\hat{n})$, where $\gamma_0$ is the surface tension of an interface
oriented in the $001$, or symmetrically equivalent direction\cite{BN}. In
experiments with Pb crystals in equilibrium with their vapor, Heyraud and
M\'etois used the inverse Wulff construction (see Ref.~\cite{BN}) to obtain
the angular part of the surface tension $\gamma(\hat{n})$ as a function of
orientation $\hat{n}$ and the same method has been employed by Surnev et
al.~\cite{SAC}. Here we use it to obtain the orientation dependence of the
surface tension for each ECS as obtained in section \ref{se:ECS}.

The surface tension $\gamma(\hat{n})$ is proportional to the distance from the
center of the cluster to the tangent plane perpendicular to $\hat{n}$,
touching the iso-surface in the point $\vec{X}$. If the ECS is scaled such
that the distance to the surface from the center of the shape along the
lattice axes is unity, then
\begin{equation} \label{invWulff}
\gamma(\hat{n}) = \max_I \hat{n} \cdot \vec{X}, 
\end{equation} 
where $\vec{X}$ is an element of the scaled iso-surface $I$. Thus, for every
direction $\hat{n}$, we have to find the point $\vec{X}$ in the iso-surface,
for which the number $\hat{n} \cdot \vec{X}$ is maximal. This procedure works
in both two and three dimensions. 

\begin{figure}
\begin{center}
\epsfxsize 7.5cm \epsfbox{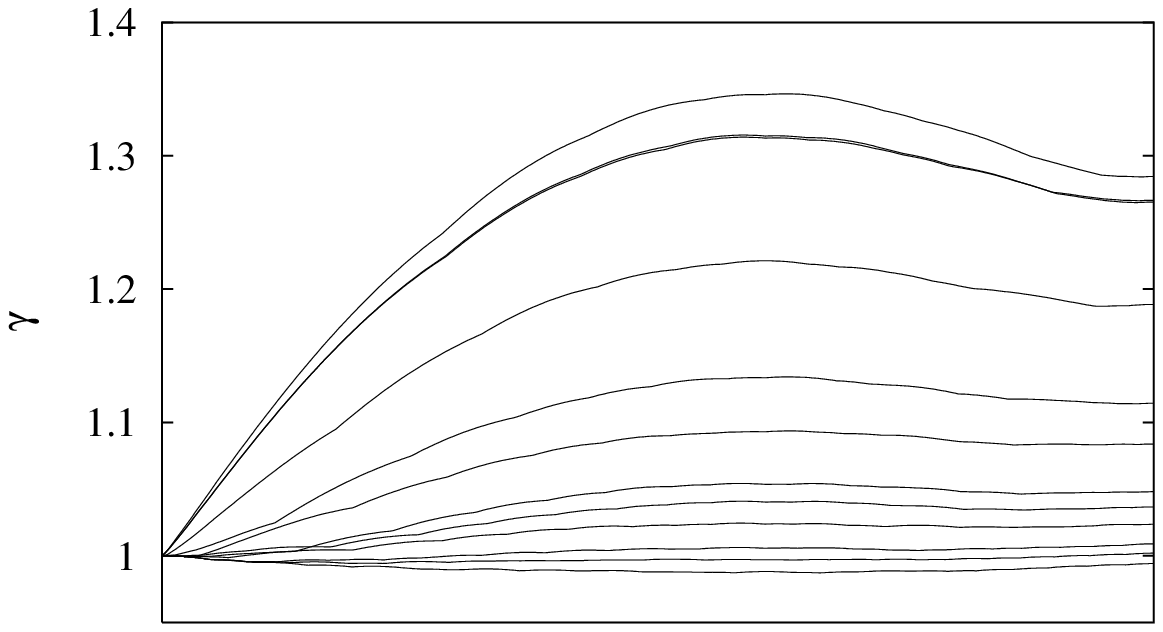}
\nopagebreak
\epsfxsize 7.5cm \epsfbox{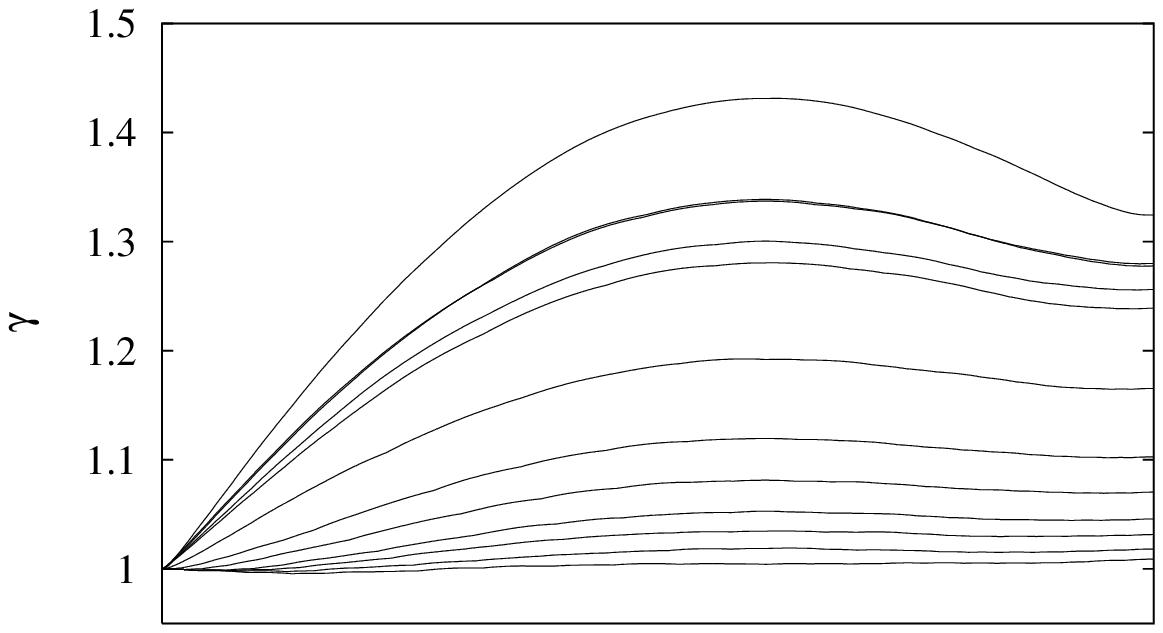}
\nopagebreak
\epsfxsize 7.5cm \epsfbox{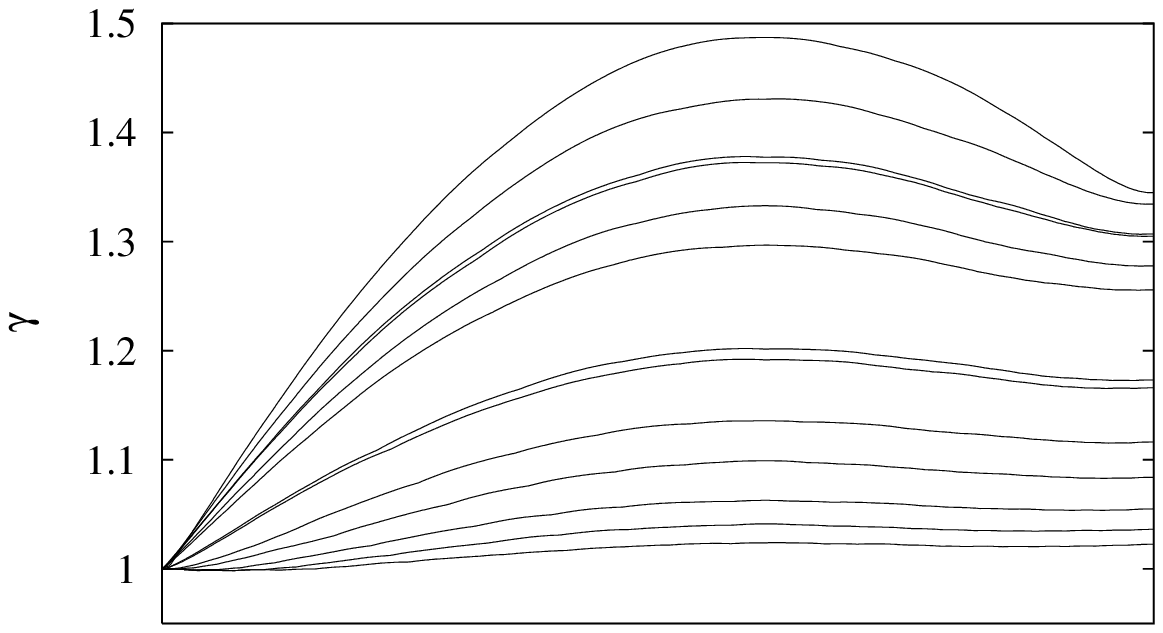}
\nopagebreak
\epsfxsize 7.5cm \epsfbox{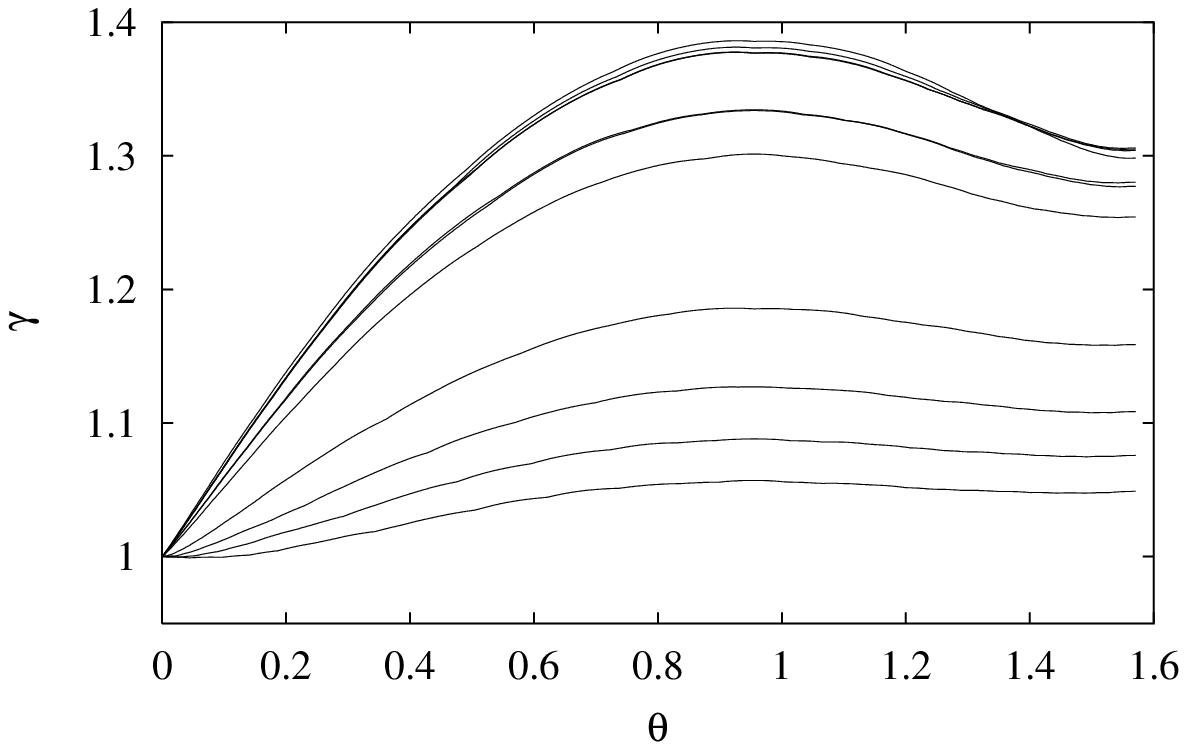}
\caption{\label{wulff}
Angular part of the surface tension $\gamma(\hat{n})$ as a function of
orientation $\hat{n}$, measured along the arc that connects the $001$ (at
$\Theta=0$), $111$ (at $\Theta=0.955$) and $110$ directions (at
$\Theta=\pi/2$, where $\theta$ is the azimuthal angle). From top to bottom the
plots show the 2, 3, 4 and 6 states results. In each plot, from the uppermost
curve downwards, the temperature ranges from 0.25 $T_m$ to 0.80 $T_m$, in
increments of 0.05 $T_m$. }
\end{center}
\end{figure}

Using ECS's obtained from fully symmetrized density profiles, we measured
$\gamma(\hat{n})$ along the arc in the $11k$-zone, which connects the $001$,
$111$ and $110$ directions. Figure~\ref{wulff} shows the results. 

These figures show that the angular dependence of the surface tension
becomes nearly constant at increasing temperature, for all numbers of
states $q$ considered here. Thus, for $q \le 6$ the ECS approaches a
sphere. For the Ising model ($q=2$) this was to be expected, since this
model undergoes a continuous phase transition at the melting point.
On the other hand, it is also clear that aspheric deviations become 
larger with increasing $q$. As we
saw in Section~\ref{se:largeq} the ECS becomes nearly cubic for large $q$.
Obviously $q=6$ still should be considered a small $q$-value in this
context.

If the ECS is not faceted, the angular part of the surface tension will
approach the point $\Theta=0$ with zero slope. If however the ECS has
$001$ facets, the approach to the point $\Theta=0$ will occur with a
non-zero slope, resulting in a cusp in the slope along the arc through
this point~\cite{BN}. Looking for the temperature where these cusps first
appear is an alternative way to measure the roughening temperature. We
found this to be less accurate than the procedure described in section
\ref{se:Tr}. The results were however consistent.

\section{Discussion and future research}
\label{Se:Discus}

We studied the Potts model with $q=2$, 3, 4, and 6 states, forced into
coexistence by fixing the density of one state.  The resulting ECS was
studied as a function of temperature and number of states.

We found that the roughening transition, which is well known for the Ising
model (equivalent to the two-states Potts model), persists for higher
numbers of states, at least up to six states. The temperature $T_R$ at
which this roughening transition occurs, measured as a fraction of the
melting temperature, tends to increase with increasing number of states.

In the future, we want to study the more general and richer behavior
of the ECS in case more than one quantity is conserved. For instance,
in the three-state Potts model close to its melting point, with the
constraint $\rho_1\gg\rho_2 \ge \rho_3$, the ECS resembles the shape of
two soap bubbles with a common interface; this changes under variations
of temperature and the ratio $\rho_2/\rho_3$.

We also want to look at equilibrium shapes in constrained geometries,
like fluids between parallel plates, or systems with grain boundaries.

Finally we are investigating the behavior for larger $q$-values with
the aid of different Monte Carlo techniques.

\end{document}